\documentclass[10pt,aps,prl,twocolumn,superscriptaddress]{revtex4}

\usepackage{epsfig}
\usepackage{dcolumn}% Align table columns on decimal point
\usepackage{bm}% bold math
\usepackage{amsmath}
\usepackage{amsfonts}
\usepackage{latexsym}
\usepackage{amssymb}
\usepackage{color}
\usepackage{hyperref}
\usepackage[normalem]{ulem}
\newcommand{\g}[1]{{\bf #1}}

\newcommand{\eqn}[1]{(\ref{#1})}
\newcommand{\be}{\begin{equation}}
\newcommand{\ee}{\end{equation}}
\newcommand{\bea}{\begin{eqnarray}}
\newcommand{\eea}{\end{eqnarray}}
\newcommand{\ba}{\begin{eqnarray*}}
\newcommand{\ea}{\end{eqnarray*}}
\newcommand{\dagga}{{\phantom{\dagger}}}

\newcommand{\U}{\text{U}}

\begin{document}
 
 \title{Interplay of charge and spin dynamics after an interaction quench\\  in the Hubbard model}
 
\author{Marcin M. Wysoki\'nski}
\email{mwysoki@sissa.it}

\affiliation{International School for Advanced Studies (SISSA)$,$ via Bonomea 265$,$ IT-34136$,$ Trieste$,$ Italy}
\affiliation{Marian Smoluchowski Institute of Physics$,$ Jagiellonian
University$,$ 
ulica prof. S. \L ojasiewicza 11$,$ PL-30-348 Krak\'ow$,$ Poland}

\author{Michele Fabrizio}
\email{fabrizio@sissa.it}
\affiliation{International School for Advanced Studies (SISSA)$,$ via Bonomea 265$,$ IT-34136$,$ Trieste$,$ Italy}

\date{\today}

\begin{abstract}
We investigate the unitary dynamics following a sudden increase $\Delta U>0$ of repulsion in the paramagnetic sector of the half-filled Hubbard model on a Bethe lattice,  by means of a variational approach that combines a Gutzwiller wavefunction with a partial Schrieffer-Wolff transformation, both defined through time-dependent variational parameters. Besides recovering at $\Delta U_c$ the known dynamical transition linked to the equilibrium Mott transition, we find pronounced
dynamical anomaly  at larger 
$\Delta U_*>\Delta U_c$  manifested in a singular behaviour of the long-time average of double occupancy. Although the real-time dynamics of the variational parameters at $\Delta U_*$ strongly resembles the one at $\Delta U_c$, careful frequency spectrum analysis suggests a dynamical crossover, instead of a dynamical transition, separating regions of a different behaviour of the spin-exchange.     
\end{abstract}

\maketitle
 
Pump-probe time-resolved spectroscopy is growing in importance as a tool for studying and manipulating  correlated materials \cite{Mihailovic2016}. On one hand it gives access to new enlightening information about the dynamical properties of those materials, beyond reach of conventional spectroscopy. 
In addition, it provides a very efficient tool to drive phase transitions on ultra-short time scales, and concurrently investigate them in the time domain. There are cases where the photo-induced phases are actually those observed at thermal equilibrium upon heating, like for instance in photo-excited VO$_2$~\cite{Cavalleri-VO2-PRL2001}. This might be suggestive of a quasi-thermal pathway, though that is not generically the case, even in the same VO$_2$~\cite{Leitenstorfer-PRB2015}. Indeed, there are evidences 
of photo-induced hidden phases that are absent at equilibrium~\cite{Mihailovic-Science2014,Averitt-NatMat2016}, as well as of remarkable non-thermal transient properties~\cite{Fausti-Science2011,Mitrano2016}.

At first sight it might seem unsurprising to observe non-thermal behaviour in correlated electron systems, which are complex materials with several competing phases and many actors playing a role. In reality, even the simplest among all models of correlated electrons, i.e. the single-band Hubbard model where the complexity of real materials is reduced just to the competition between the on-site repulsion $U$ and the nearest-neighbour hopping $t$, shows puzzling and still controversial non-thermal behaviour. In a seminal work~\cite{Kollar2009}, 
Eckstein, Kollar and Werner discovered by time-dependent dynamical mean-field theory (t-DMFT) that the unitary evolution of the half-filled Hubbard model in an infinitely coordinated Bethe lattice after a sudden increase of the repulsion $\text{U}$ from the initial $\U_0=0$ to a final $\U_f>0$ displays a sharp crossover at $\U_f=\U_c$, which, within the relatively-short numerically-affordable simulation times,   resembles a genuine dynamical transition. Such an interpretation was however contrasted by the observation that, if one assumes thermalisation,  the temperature that corresponds to the energy supplied by the quench $\U=0\to \U_c$ is well above the second-order critical endpoint of the Mott-transition line in the $T$ vs. $\U$ equilibrium phase diagram~\cite{Kollar2009}. 
The same dynamical crossover was later found, still in an infinitely coordinated lattice, by a variational approach based on a time-dependent Gutzwiller wavefunction~\cite{Schiro2010,Schiro2011}, 
which is not as rigorous as t-DMFT but allows simulating much longer times. In particular, by considering a linear ramp rather than a sudden quench, this crossover was shown~\cite{Sandri2012} to be a genuine dynamical transition linked to the equilibrium Mott transition. The same conclusion has been recently drawn by the nonequilibrium self-energy functional theory~\cite{Potthoff2016}, which is supposed to be more rigorous than the variational Gutzwiller approach, though not as much as t-DMFT. We emphasise that, should this dynamical anomaly be confirmed to correspond to a dynamical Mott transition, it would imply~\cite{Kollar2009} that even the simplest single-band Hubbard model may display non-thermal behaviour, at least in lattices with infinite coordination number. However, there are evidences that the same occurs also when the coordination is finite~\cite{Hamerla2013,Hamerla2014}.  

It is therefore worth proving or disproving the existence of this dynamical transition by other complementary techniques, looking forward to numerical developments that could allow t-DMFT to finally settle this issue.    
Here we make such an attempt by extending out of equilibrium the variational approach that we recently proposed~\cite{Wysokinski2017R}, and which combines a Gutzwiller wavefunction with a partial Schrieffer-Wolff transformation, both defined now by time-dependent variational parameters. At equilibrium, this variational wavefunction provides a much better description of the Mott insulator than the simple Gutzwiller wavefunction, and it is therefore likely it may also describe more accurately the non-equilibrium dynamics.  
 
We consider the half-filled single-band Hubbard model on an infinitely coordinated Bethe lattice,
\begin{equation}
H(t)= -\frac{ 1}{\sqrt{z}}\,\sum_{<ij>}T_{\g i\g j}  
+ \frac{{\rm U}(t)}{2}\sum_{\g  i}\big(n_{\g i\uparrow}+n_{\g i\downarrow}-1\big)^2, 
\label{HM}
\end{equation}
where $z\to\infty$ is the coordination number, ${\rm U}(t)$ the time-dependent on-site repulsion, ${\rm U}(t<0)={\rm U}_0$ and ${\rm U}(t\geq0)={\rm U}_{f}$, and 
${T_{\g i\g j}\equiv\sum_\sigma (c_{\g i\sigma}^\dagger c_{\g j\sigma}^\dagga+ 
H.c.)}$ the hopping operator on the bond $<\!ij\!>$ connecting 
the nearest-neighbour sites $i$ and $j$.

The dynamics of the model is determined through the saddle point of the action 
\begin{equation}
  \mathcal{S}=\int dt \;\langle\Psi(t)|\;i\frac{d}{d t}-H(t)\;|\Psi(t)\rangle\, ,
\label{lag}
\end{equation}
which provides the exact solution of the Schr\"odinger equation for unrestricted many-body wavefunctions 
$\Psi(t)$, or just a variational estimate of it in case $\Psi(t)$ varies within a subspace of the whole Hilbert space, which is what we shall do hereafter. In particular, we assume for $\Psi(t)$ the expression \cite{Wysokinski2017R},
\begin{equation}
  \big|\Psi(t)\big\rangle = \mathcal{U}(t)\,\mathcal{P}_G(t)\,\big|\psi_0(t)\big\rangle\,, \label{psi}
 \end{equation}
where $\psi_0(t)$ is a paramagnetic uniform Slater determinant, $\mathcal{U}(t)$ a unitary transformation, and finally $\mathcal{P}_G(t)\equiv\prod_{\g i} \mathcal{P}_{ \g i}(t)$, with $\mathcal{P}_{ \g i}(t)$ a linear operator on the local Hilbert space \cite{Schiro2010}. In the presence of particle-hole symmetry and discarding spontaneous breakdown of 
spin $SU(2)$ symmetry, $\mathcal{P}_{ \g i}(t)$ can be generically written as 
\begin{equation}
\begin{split}
&\!  \mathcal{P}_{ \g i}(t)\!=\! \sqrt{2} \;\phi_{\g i0}(t)\Big[{\rm P}_{\g i}(0)\!+\!{\rm P}_{\g i}(2)\Big]+\sqrt{2}\; \phi_{\g i1}(t) \,{\rm P}_{\g i}(1) \,,  
\end{split}
\end{equation}
where ${\rm P}_{\g i}(n)$ is the projection operator at site $i$ onto the configuration with $n$ electrons, whereas $\phi_{\g in}(t)$ is a complex function of time~\cite{Schiro2010}. In infinitely coordinated lattices the wavefunction $|\Psi(t)\rangle$ is normalised at any time if ${|\phi_{\g i0}(t)|^2+|\phi_{\g i1}(t)|^2=1}$.

The time-dependent unitary transformation $\mathcal{U}(t)$ is of the Schrieffer--Wolff type~\cite{Spalek1977,Eckstein2017,Wysokinski2017R}, and it is parametrised by complex, time and bond dependent variational parameters $\epsilon_{\g i \g j}(t)$:   
\begin{equation}
 \mathcal U(t) \equiv{\rm e}^{A(t)}\equiv\exp\Bigg[\frac{1}{\sqrt{z}} \sum_{\langle\g i\g j\rangle}\Big(\epsilon_{\g i\g j}(t)\,\tilde T^\dagga_{\g i\g j} -\epsilon^*_{\g i\g j}(t)\,\tilde T^\dagger_{\g i\g j}\,\Big)\Bigg],
\label{U}
\end{equation}
where
\begin{equation}
\begin{split}
\! \tilde  T_{\g i\g j}&\equiv  \Big({\rm P}_{\g i}(2){\rm P}_{\g j}(0)+ {\rm P}_{\g i}(0){\rm P}_{\g j}(2)\Big)\,T_{\g i\g j}\,\Big({\rm P}_{\g i}(1){\rm P}_{\g j}(1)\Big),\\
\! \tilde  T_{\g i\g j}^\dagger&\equiv\Big({\rm P}_{\g i}(1){\rm P}_{\g j}(1)\Big)\,T_{\g i\g j}\,\Big({\rm P}_{\g i}(2){\rm P}_{\g j}(0)+{\rm P}_{\g i}(0){\rm P}_{\g j}(2)\Big),
\end{split}
\end{equation} 
are the components of the hopping operator $T_\text{ij}$ that couple the low-energy subspace of singly occupied sites $i$ and $j$ with the high-energy one where one site is empty and the other doubly occupied.  
 
We determine $\psi_0(t)$, $\mathcal{P}_G(t)$ and $\mathcal U(t)$ through the saddle point of the action \eqref{lag} with respect to all the variational parameters, handling $\mathcal U(t)$ by a series expansion 
\begin{equation}
  U^\dagger \mathcal{O}\,U \!=\!   \mathcal{O}- [A, \mathcal{O}]
  +\frac{1}{2}\,\Big[A,[A,\mathcal{O}]\Big] +\dots\,,
\label{expa}
\end{equation}
up to the desired order. For instance 
\begin{equation}
\begin{split}
U^\dagger\, H\,U^\dagga\!\!&\simeq H+\frac{U}{\sqrt{z}}\,\sum_{\g i \g j}\Big(\epsilon_{\g i \g j} \,\tilde T_{\g i \g j}+\epsilon^*_{\g i \g j} \,\tilde T^\dagger_{\g i \g j}\,\Big) \\
&+\frac{1}{8z}\sum_{\g i \g j} J_{\g i \g j}\, \Big[\tilde T_{\g i \g j}+\tilde T^\dagger_{\g i \g j}\, ,\, \tilde T_{\g i \g j}-\tilde T^\dagger_{\g i \g j}\,\Big]+ H_{\mathcal R}\,,
\label{TH}
\end{split}
\end{equation}
where,
\begin{equation}
J_{\g i \g j}=4\,\Big[\big(\epsilon_{\g i \g j}+\epsilon^*_{\g i \g j}\big)\,V - U\, \big|\epsilon_{\g i \g j}\big|^2\,\Big]\,,
\label{Jj} 
\end{equation}
and higher order terms are stored together in $H_{\mathcal R}$. In the calculation below we stop the series expansion at the third order in power of $\epsilon$, and consider all processes up to three neighbouring sites. We have tested such an approximation at equilibrium in comparison with exact DMFT results \cite{Adriano}, and it provides a quite satisfactory description of metal and insulating phases for $\U\gtrsim \U_{\rm Mott}/2$, where $\U_{\rm Mott}$ is the equilibrium location of the Mott transition (cf. Supplemental Material, Sec.I \cite{SM}). Inclusion of higher orders systematically increases the accuracy and thus allows accessing also the weaker correlated regime~\cite{Wysokinski2017R}. However, for the sake of simplicity, we decided to stand to the above approximation, and consequently we just considered quantum quenches from a relatively correlated metal at $\U_0> \U_{\rm Mott}/2$ to higher values of $\U_f>\U_0$. 

With this prescription for handling the unitary operator $\mathcal{U}(t)$,  the expectation values that define the action \eqref{lag} can be explicitly evaluated when the coordination number $z\to\infty$, and can be 
formally written as 
\begin{equation}
\begin{split}
 \langle \Psi|i\frac{d}{d t}|\Psi\rangle&=  i\,\langle \psi_0|\dot\psi_0\rangle+i\sum_{\g i}\Big(\phi_{\g i 1}^*\,\dot\phi_{\g i 1}^\dagga+\phi_{\g i 0}^*\,\dot\phi_{\g i 0}^\dagga\Big)\\ 
&+i\, f\Big(\mathbf{v},\dot \epsilon_{\g i\g j},\dot{\epsilon}_{\g i\g j}^*,\psi_0,\psi_0^*\Big)\,,\\
\langle \Psi|H|\Psi\rangle&=  h\big(  \mathbf{v},\psi_0,\psi_0^*\big),
\label{Lr}
\end{split}
\end{equation}
where $\mathbf{v} =\{\phi^\dagga_{\g i 0},\phi^*_{\g i 0},\phi_{\g i 1}^\dagga,\phi_{\g i 1}^*,\epsilon_{\g i\g j}^\dagga,\epsilon_{\g i\g j}^*\}$. Being too lengthy,  
the actual expressions of the functions $f$ and $h$ are given in the Supplemental 
Material, Sec.II \cite{SM}. 

The saddle point equations that determine the evolution of the wavefunction can be readily obtained. Like in the time-dependent Gutzwiller approximation \cite{Schiro2010}, the evolution of the Slater determinant 
$\psi_0(t)$ is trivially just the multiplication by a time dependent phase, so that, for instance, 
${\langle\psi_0(t)|\frac{1}{\sqrt{z}}\sum_{ij}T_{\g i\g j}|\psi_0(t)\rangle=8/3\pi\equiv T_0}$ is time independent. In what follows, we shall use as energy unit $8T_0$, and define $u\!=\!{\rm U}/8T_0$. In these units, the initial state is always prepared at $u_0=0.5$ and then instantly quenched to the final value $u_f>0.5$.

\begin{figure}[thb]
 \centerline{\includegraphics[width=0.47\textwidth]{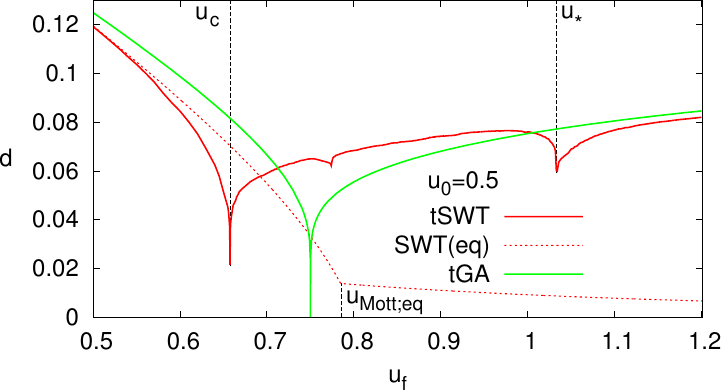} }
\caption{Average double occupancy after the quench from the correlated metal at $u_0=0.5$. For the sake of completeness, we also show the equilibrium result (SWT) \cite{Wysokinski2017R} as well as that obtained by the simpler Gutzwiller wavefunction (tGA)~\cite{Schiro2010}.}
\label{fig1}
\end{figure}

On the contrary, the Euler-Lagrange equations for the components of the variational parameter $\mathbf{v}$ 
are not trivial and read
\begin{equation}
\begin{split}
 i\dot\phi_{\g i 0}^\dagga+i\frac{\partial f}{\partial \phi_{\g i 0}^*}-\frac{\partial h}{\partial \phi_{\g i 0}^*}&=0\,,\\
i\dot\phi_{\g i 1}^\dagga+i\frac{\partial f}{\partial \phi_{\g i 1}^*}-\frac{\partial h}{\partial \phi_{\g i 1}^*}&=0\,,\\
i\frac{\partial f}{\partial \epsilon_{\g i \g j}}-\frac{\partial h}{\partial \epsilon_{\g i \g j}}-i\frac{d}{dt}\frac{\partial f}{\partial \dot \epsilon_{\g i \g j}}&=0\,,
\end{split}\label{Euler-Lagrange}
\end{equation}
plus their complex conjugates. 
Assuming translational invariance, we can discard the site index in the above equations. 
The resulting differential equations are lengthy but can be written in the following matrix form  
\begin{equation}             
\hat B\big[\mathbf{v}(t)\big] \; \dot{\mathbf{v}}(t) = \mathbf{a}\big [\mathbf{v}(t)\big]\,,                                                                                                                                                               
\label{RK}
\end{equation}
i.e. like a set of ordinary 
first order non-linear differential equations, which can be numerically integrated by Runge-Kutta type of algorithms, see Supplemental Material, Sec.III \cite{SM}. 

\begin{figure}[t]
\centerline{\includegraphics[width=0.5\textwidth]{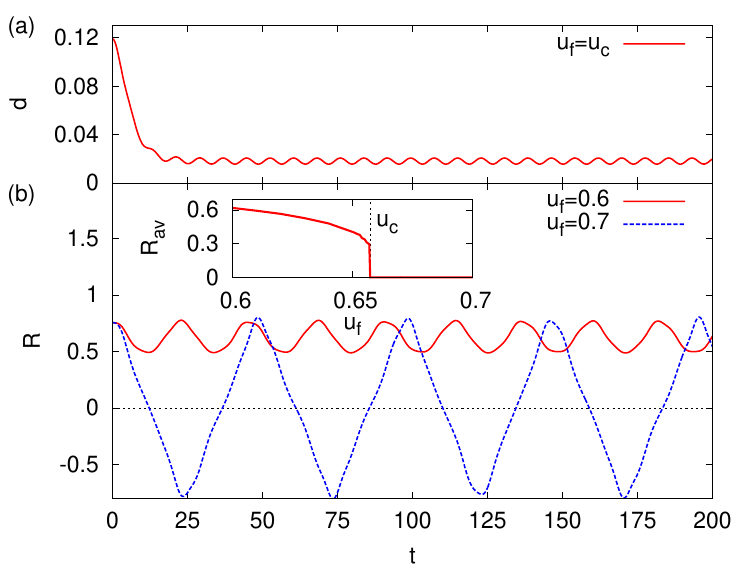}} 
\caption{Panel (a): time evolution of the double occupancy $d(t)$ at $u_c$. Panel (b): time evolution of 
$R(t)$ just before and after $u_c$, with its time average drawn in the inset. We note that indeed $R$ has a critical behaviour at $u_c$.}
\label{fig2}
\end{figure}

\begin{figure*}[t]
\centerline{ \includegraphics[width=0.49\textwidth, trim={0  0 0 0.2cm },clip]{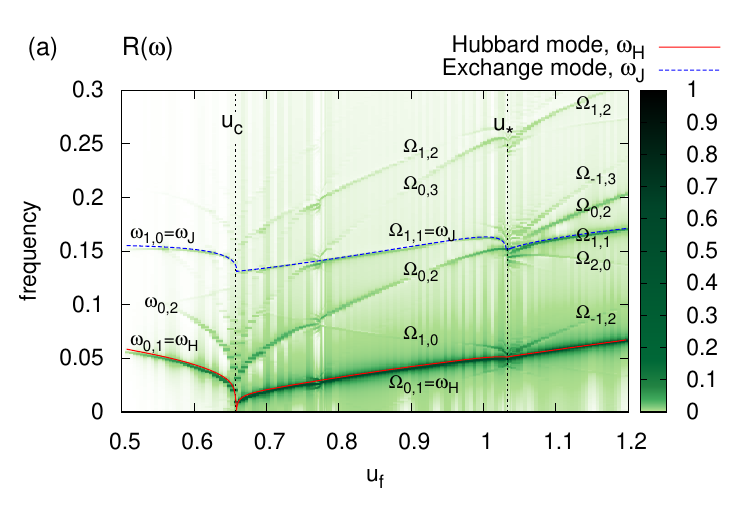} 
 \includegraphics[width=0.49\textwidth, trim={0 0 0 0.2cm },clip]{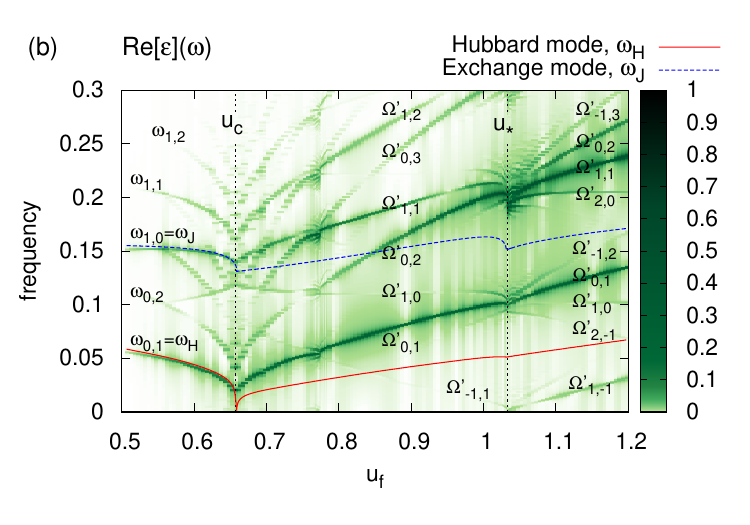} }
%\vspace{-0.7cm} 
\caption{Frequency spectra of $R(t)$, i.e. the real part of $\phi_0\phi_1^*$, and of the real part of $\epsilon$, 
panels (a) and (b), respectively. The meaning of the frequency labels is explained in the main text.}
\label{fig3}
\end{figure*}

In Fig. \ref{fig1} we plot the long-time average of the double occupancy per site. At $u_{c}\simeq 0.6575$ we observe a first dynamical anomaly, which is actually the already known {\it dynamical transition} \cite{Kollar2009,Schiro2010,Potthoff2016} at which the system shows a rapid relaxation to a Mott insulator (cf. Fig. \ref{fig2}a). Within Gutzwiller type of wavefunctions, the Mott transition is characterised by an order parameter $R=\phi_0\,\phi_1^*+ \phi_1\,\phi_0^*$, which is finite in the metal and vanishes in the insulator~\cite{Schiro2010,Zitko-PRB2015}. Formally $R$ is defined by observing that the action of the 
projected operator $\mathcal{P}_i^\dagger \,c^\dagga_{i\sigma}\,\mathcal{P}_i$ on the Slater determinant 
$\psi_0$ is the same as $R\, c^\dagga_{i\sigma}$, so that $R$ can be regarded as the quasiparticle component in the physical electron $c^\dagga_{i\sigma}$. In Fig.~\ref{fig2}b we show that, for $u_f<u_c$, $R(t)$ oscillates 
around a finite value, whereas above $u_c$ it oscillates around zero, as clear in the inset where its time average is ploted. Therefore our improved wavefunction also points to a genuine dynamical Mott transition occurring at $u_c$, which, as we mentioned, contrasts the belief that the system thermalises. 

In addition, we note two further dynamical anomalies at 
$u_f\simeq 0.77$ and $u_f=u_{*}\simeq1.0326$, the latter more pronounced. To better explore their nature, in Fig.~\ref{fig3} we draw the frequency spectra of $R(t)$  and of the real part of $\epsilon(t)$, which 
show that both anomalies are actually 
triggered by the frequency crossing of different modes. In addition, at $u_*$ there is also one mode that gets soft, not much different from what happens at $u_c$. In order to identify the origin of the different modes and the meaning of the softening, we compare the frequency spectrum of $R(t)$ with the corresponding one in the simpler time-dependent Gutzwiller wavefunction~\cite{Schiro2010}, see Sec.IV of the Supplemental Material \cite{SM}, which lacks the spin correlations brought by the Schrieffer-Wolff transformation.
Both spectra have in common the mode with frequency $\omega_H$, see Fig. \ref{fig3}a. We thus conclude 
that $\omega_H$ originates solely from the dynamics of charge degrees of freedom, and can be 
associated with the \textit{Hubbard-band mode} discussed in Ref.~\cite{Fabrizio-PRB2017} that becomes soft at the equilibrium Mott transition.  In the present non-equilibrium condition, the softening of the same  mode and its higher harmonics is a further confirmation that $u_c$ signals a genuine dynamical Mott transition. 

On the other hand, the Schrieffer-Wolff transformation leads to the appearance of a new mode that does not softens at $u_c$, and which we associate to the spin exchange 
$J$ of Eq.~\eqn{Jj} and thus denote as exchange mode $\omega_J$, marked 
with a blue dashed line in Fig.~\ref{fig3}. 

\begin{figure}[bht]
 \centerline{\includegraphics[width=0.49\textwidth]{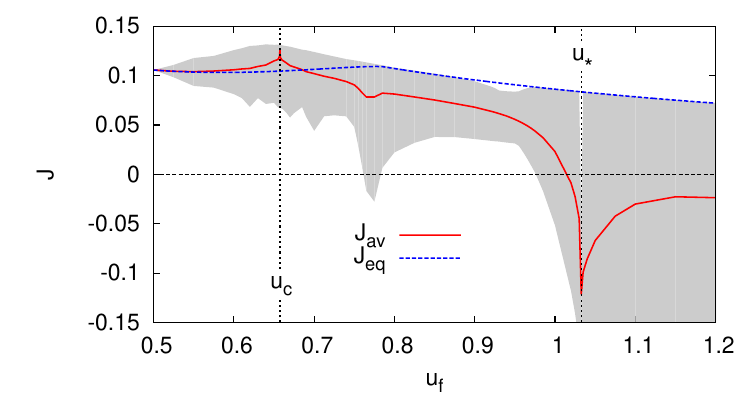} }
%\vspace{-0.9cm} 
\caption{Red solid line: time average value $J_\text{av}$ of $J(t)$ in Eq.~\eqn{Jj}, as function of $u_f$. Blue dashed line: the value $J_\text{eq}$ of $J$ obtained at equilibrium by minimising the energy at $u=u_f$. In grey we indicate the region covered by the time fluctuations of $J(t)$. }
\label{fig4}
\end{figure}

We have found that all remaining frequencies in Fig.\ref{fig3}a are quantised linear combinations of the two principal frequencies $\omega_H$ and $\omega_J$, as expected because of the non-linear character of the Euler-Lagrange equations \eqn{Euler-Lagrange}. We shall label those secondary frequencies   
by two integers $n,k$, and denote them as $\omega_{n,k}$ for $u_f<u_c$, while, above $u_c$, 
as $\Omega_{n,k}$ and $\Omega'_{n,k}$, in the power spectra of $R(t)$, Fig.~\ref{fig3}a, and 
$\epsilon(t)$, Fig.~\ref{fig3}b, respectively. These frequencies are constructed according to the following rules:
\begin{equation}
\begin{split}
 \omega_{n,k}&=n\,\omega_J+k\,\omega_H\,,\\
\Omega_{n,k}&=n\,\big(\omega_J-\omega_H\big)+(2k-1)\,\omega_H\,,\\
\Omega'_{n,k}&=\Omega_{n,k}+\omega_H=n\,(\omega_J-\omega_H)+2k\,\omega_H\,.
\label{emp}
\end{split}
\end{equation}
Accordingly, mode crossings occur whenever $\omega_J=(2m+1)\,\omega_H$, and thus the anomaly at 
$u_f\simeq 0.77$ corresponds to $m=2$ while that at $u_*$ to $m=1$. Moreover,  
the apparent softening in the dynamics of $\epsilon$ is in reality the vanishing of the linear combination $\Omega'_{\pm 1,\mp 1}$. In other words, the anomaly at $u_*$ is not characterised by the softening of any of the principal modes, $\omega_H$ or $\omega_J$, and therefore it is not to be confused with a genuine dynamical transition. Nonetheless, close to $u_*$ we do find changes in physical properties. 
In Fig.~\ref{fig4} we plot as function of $u_f$ 
the time average $J_\text{av}$ of the spin exchange $J(t)$ in Eq.~\eqn{Jj}, the region covered by its time fluctuations, as well as its equilibrium value $J_\text{eq}$. We observe that, just before $u_*$, $J_\text{av}$ 
turns from antiferromagnetic to ferromagnetic and its fluctuations grow larger, which suggests a change in character of the spin correlations. We argue it might correspond to the melting of antiferromagnetism that 
should occur when suddenly increasing $\U$ starting from a N\'eel ordered state~\cite{Sandri2013}. 
 
% \vspace{-0.9cm} 

In summary, we have studied the quench dynamics in the paramagnetic sector 
of the half-filled single-band Hubbard model on an infinitely coordinated Bethe-lattice, by means of a 
variational Gutzwiller wavefunction enriched with spin correlations by a variational Schrieffer-Wolff transformation. We have confirmed the existence of a dynamical Mott transition at odds with the belief that 
the model should finally relax to a thermal state. Even though the variational wavefunction does not allow for all dissipative channels that exist in the real time evolution, we nonetheless believe that the softening of the Hubbard-band mode with frequency $\omega_H$, which signals the Mott transition at equilibrium~\cite{Fabrizio-PRB2017}, is a genuine phenomenon that will not be 
swept out in more rigorous calculations.  
In addition we have found that the time-dependent Schrieffer-Wolff transformation yields non-trivial spin correlations that undergo a dynamical change for a final interaction value quite beyond the dynamical Mott transition.

{\it Acknowledgements.} MMW acknowledges support 
from the Polish Ministry of Science and Higher Education under 
the ``Mobility Plus'' programme, Agreement No. 1265/MOB/IV/2015/0, as well as from the Foundation for Polish Science under the ``START'' programme.
 MF acknowledges   support from  European Union under the H2020 Framework Programme, ERC Advanced  Grant  No.   692670  ``FIRSTORM".  
%  \bibliography{SW.bib}

\widetext
\clearpage

\begin{center}
{\Large \bf Supplemental Material  to   ``Interplay of charge and spin dynamics\\ after an interaction quench  in the Hubbard model''} 
\end{center}
\endwidetext

\section{Accuracy of expansion at equilibrium}
\begin{center}
\begin{figure}[b]
 \includegraphics[width=0.5\textwidth]{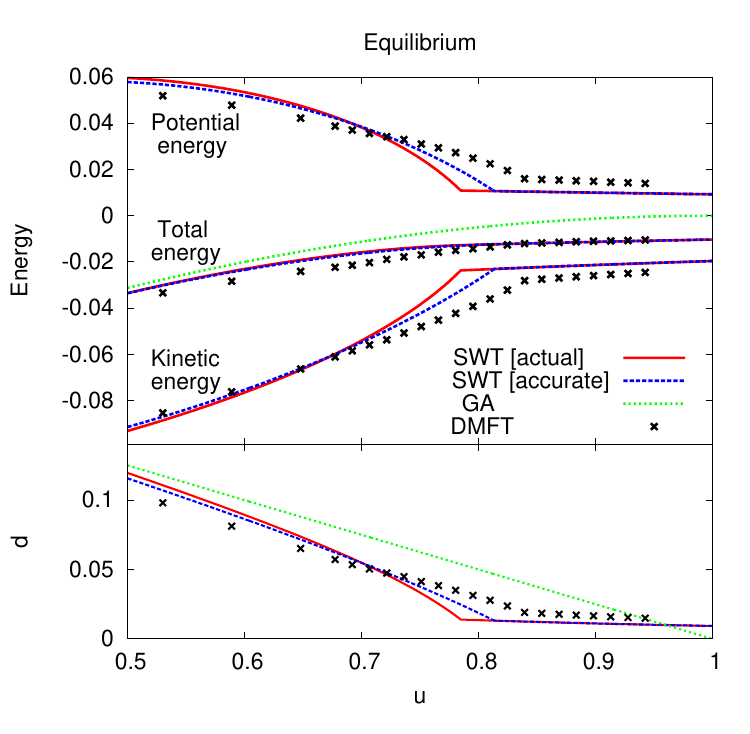} 
\caption{ Comparison between equilibrium energies and averaged double occupancies ($d$) between different many-body methods: SWT [actual] - present technique under actual approximation;  SWT [accurate] - present technique under more accurate approximation incorporating incoherent processes involving 4 sites (used in Ref. \onlinecite{Wysokinski2017RS}); GA - Gutzwiller approximation, or 0th order expansion in power of $\epsilon$ of the present technique; DMFT - dynamical mean field theory providing numerically exact solution \cite{AdrianoS}.}
\label{sfig1}
\end{figure}
\end{center}
In our present considerations, while expanding unitary transformed operators, we stop expansion at the third order in power of $\epsilon$ and we have taken into account all incoherent processes involving two and three neighbouring sites on the Bethe lattice. This approximation provides a satisfactory  description of the paramagnetic state of the Hubbard model for $U/8T_0\gtrsim0.5$ when compared to the exact results of dynamical mean-field theory (DMFT) \cite{AdrianoS} (cf. Fig. \ref{sfig1}). 
For the sake of completeness we have also plotted results obtained by extending our method by including also four site processes that already properly describes the system at any $U$ \cite{Wysokinski2017RS}.
The departures from the more accurate approximation for $U/8T_0\gtrsim0.5$ are small and only  quantitative.

%  \vspace{-0.7cm}

\section{Explicit form of functions determining action}
For the sake of completeness we explicitly show how to approximate time derivative operator by taking its average with the unitary transformed Gutzwiller wavefunction,
\begin{equation}
\begin{split}
 \langle \psi_0|\mathcal{P}_G^\dagger  &U^\dagger i\frac{d}{d t}U\mathcal{P}_G^\dagga|\psi_0\rangle\simeq\ i\langle \psi_0|\dot\psi_0\rangle+i\langle \psi_0|\mathcal{P}_G^\dagger\dot{\mathcal{P}}_G^\dagga|\psi_0\rangle\\
&+i\langle \psi_0|\mathcal{P}_G^\dagger\dot A\mathcal{P}_G^\dagga|\psi_0\rangle- \frac{i}{2}\langle \psi_0|\mathcal{P}_G^\dagger [A,\dot A]\mathcal{P}_G^\dagga|\psi_0\rangle\\
&+ \frac{i}{6}\langle \psi_0|\mathcal{P}_G^\dagger \big[A,[A,\dot A]\big]\mathcal{P}_G^\dagga|\psi_0\rangle, 
\label{dtS} 
\end{split}
\end{equation}
where,
\begin{equation}
 \begin{split}
&  i\langle \psi_0|\mathcal{P}_G^\dagger\dot{\mathcal{P}}_G^\dagga|\psi_0\rangle=i\sum_{\g i}(\phi_{1\g i}^*\dot\phi_{1\g i}+\phi_{0\g i}^*\dot\phi_{0\g i}).
\end{split}
\end{equation}

Now we can readily provide explicit  forms of functions $f$ and $h$ introduced in the main manuscript ($N=\sum_{\g i}1$),
\begin{equation}
 \begin{split}
  \frac{f}{N}=&\frac{T_0 t}{V}(\dot \epsilon \phi_0^{*2} \phi_1^2-\dot \epsilon^* \phi_0^2 \phi_1^{*2} ) -\frac{1}{4} (\dot \epsilon \epsilon^*-\epsilon \dot \epsilon^*) (|\phi_0|^4 -|\phi_1|^4)\\
  -&\frac{3T_0^2}{V^2}(\dot \epsilon \epsilon^*-\epsilon \dot \epsilon^*)    (|\phi_0|^2 -|\phi_1|^2)|\phi_0|^2|\phi_1|^2 \\
 +&\frac{T_0}{V}(\epsilon \dot \epsilon^*-\dot \epsilon \epsilon^*)  (|\phi_0|^2 +|\phi_1|^2) \left(\epsilon \phi_0^{*2} \phi_1^2+\epsilon^* \phi_0^2 \phi_1^{*2}\right),
 \end{split}
\end{equation}
\widetext
\begin{equation} 
 \begin{split}
 \frac{h}{N}=& \frac{U}{2} \phi_0 \phi_0^* - T_0  (\phi_0 \phi_1^*+\phi_0^* \phi_1)^2+\frac{T_0 U}{V} (\epsilon \phi_0^{*2}\phi_1^2+\epsilon^* \phi_0^2 \phi_1^{*2}) +\big[\frac{ V }{2}(\epsilon+\epsilon^*)-\frac{1}{2} |\epsilon|^2 U\big] (|\phi_0|^4 -|\phi_1|^4 )\\
&+\frac{3T_0^2}{V^2}   \big[|\epsilon |^2 U+(\epsilon^*+\epsilon) V\big]
   (|\phi_0|^2-|\phi_1|^2 )|\phi_0|^2 |\phi_1|^2 
+\frac{3 T_0^2}{V}  (|\phi_0|^2-|\phi_1|^2 ) \left(\epsilon \phi_0^{*2}
   \phi_1^2+\epsilon^* \phi_0^2 \phi_1^{*2}\right)\\
&+\frac{T_0}{V}\Big[\frac{3V}{2}  (\epsilon+\epsilon^*)-|\epsilon|^2 U\Big] (|\phi_0|^2+|\phi_1|^2 ) \left(\epsilon
   \phi_0^{*2} \phi_1^2+\epsilon^* \phi_0^2 \phi_1^{*2}\right)
+3T_0  |\epsilon|^2  (|\phi_0|^2+|\phi_1|^2 )|\phi_0|^2 |\phi_1|^2\\
& -\frac{2}{3} |\epsilon|^2 V
   (\epsilon +\epsilon^* ) (|\phi_0|^4-|\phi_1|^4) (|\phi_0|^2+|\phi_1|^2).
% 
% \frac{f}{N}=&\frac{T_0 t}{V}(\dot \epsilon \phi_0^{*2} \phi_1^2-\dot \epsilon^* \phi_0^2 \phi_1^{*2} ) -\frac{1}{4} (\dot \epsilon \epsilon^*-\epsilon \dot \epsilon^*) (|\phi_0|^4 -|\phi_1|^4)  -\frac{3T_0^2}{V^2}(\dot \epsilon \epsilon^*-\epsilon \dot \epsilon^*)    (|\phi_0|^2 -|\phi_1|^2)|\phi_0|^2|\phi_1|^2 \\
%  &+\frac{T_0}{V}(\epsilon \dot \epsilon^*-\dot \epsilon \epsilon^*)  (|\phi_0|^2 +|\phi_1|^2) \left(\epsilon \phi_0^{*2} \phi_1^2+\epsilon^* \phi_0^2 \phi_1^{*2}\right).
 \end{split}
\end{equation}
\endwidetext

\section{Managing equations for Runge-Kutta algorithm}
We need to deal with a matrix ordinary differential equation in the form
\begin{equation}
 \hat B[\bf v] \dot{{\bf v}} = {\bf a} ( {\bf v}),
\end{equation}
where ${\bf v} \in\{\phi_0,\phi_0^*,\phi_1,\phi_1^*,\epsilon,\epsilon^*\}$ is a vector in which we have stored all variational variables. Moreover $\hat B$ is 6x6 matrix and ${\bf a}$ is 6 component vector, both with elements functionally depending on the variables from $\bf  v$. In order to efficiently proceed with numerical Runge-Kutta method we should put our equation in a following form 
 \begin{equation}
  \dot{{\bf v}} = \hat B^{-1}{\bf a} 
\end{equation}
Because matrix $\hat B$ is highly complicated function of variables from vector $\bf v$ it is not particularly straightforward to calculate its inverse. For that reason it is quite convenient to first  analytically reorganise equations into    manageable form. 

Usual Runge-Kutta algorithms demand functional form of the coefficients up to the second order expansion in powers of small time interval $\delta t$
% \widetext
 \begin{equation}
\begin{split}
  {\bf v}_k(t+\delta t) \simeq &\, {\bf v}_k(t)+ (\hat B^{-1}{{\bf a}})_k \delta t\\
& + \sum_i \frac{\partial (\hat B^{-1}{{\bf a}})_k}{\partial {\bf v}_i}(\hat B^{-1}{{\bf a}})_i \frac{\delta t^2}{2},
\label{srkS}
\end{split}
\end{equation}
where subscript $k$ (or $i$) denotes $k$-th (or $i$-th) element of the vector.
% \endwidetext
In above we have assumed that neither $\hat B$ nor ${\bf a}$ are explicitly time-dependent as it is the case in considered situation. 

Analytical forms of coefficients in expansion in \eqref{srkS} are generally inaccessible. For that reason in a following we provide simple  and efficient numerical prescription for evaluating these coefficients at any instance of time.   
By using the so-called LU decomposition we may easily solve following matrix equations,
\begin{equation}
 \begin{split}
& \hat B {\bf x}=  {\bf a}  \ \ \Rightarrow \ \ {\bf x}_{k}=(\hat B^{-1}{\bf a})_k,\\
& \hat B {\bf y}^i=\frac{\partial {\bf a}}{\partial {\bf v}_i} \ \ \Rightarrow \ \ {\bf y}^i_{k}= (\hat B^{-1}\frac{\partial {\bf a}}{\partial {\bf v}_i})_k, \\
&\hat B{\bf z}^i=-\frac{\partial\hat B}{\partial {\bf v}_i}{\bf x} \ \ \Rightarrow \ \  {\bf z}^i_{k}=-(\hat B^{-1}\frac{\partial\hat B}{\partial {\bf v}_i}{\bf x})_k.
\label{sxS}
 \end{split}
\end{equation}
% Note that in above equations all vector elements on the right hand side of the above equations can be explicitly obtained in a direct analytical form. 
Now taking advantage of the relation
\begin{equation}
 \frac{\partial \hat B^{-1}}{\partial {\bf v}_i}=-B^{-1}\frac{\partial \hat B}{\partial {\bf v}_i}B^{-1}
\end{equation}
we may determine 
\begin{equation}
\begin{split}
 \frac{\partial (\hat B^{-1}{\bf a})_k}{\partial {\bf v}_i}&=(\frac{\partial \hat B^{-1}{\bf a}}{\partial {\bf v}_i})_k =(\frac{\partial \hat B^{-1}}{\partial {\bf v}_i}{\bf a})_k+(\hat B^{-1}\frac{\partial {\bf a}}{\partial {\bf v}_i})_k \\&=-(\hat B^{-1}\frac{\partial \hat B}{\partial {\bf v}_i}\hat B^{-1}{\bf a})_k+{\bf y}^i_{k}\\
&=-(B^{-1}\frac{\partial \hat B}{\partial {\bf v}_i}{\bf x})_k+{\bf y}^i_{k} = {\bf z}^i_{k}+{\bf y}^i_{k}
\end{split}
\end{equation}

Finally necessary formula \eqref{srkS} for the Runge-Kutta algorithm can be expressed as,
 \begin{equation}
\begin{split}
  {\bf v}_k(t+\delta t) \simeq\ & {\bf v}_k(t)+ \delta t\,{\bf x}_{k}(t)\\
 &+ \frac{\delta t^2}{2}\sum_i [{\bf z}^i_{k}(t)+{\bf y}^i_{k}(t)]\,{\bf x}_{i}(t), 
\end{split}
\end{equation} 
% \endwidetext
which together with Eqs. \eqref{sxS} are easily manageable numerically for arbitrary long at complicated functional forms of elements of matrix $\hat B$ and vector ${\bf a}$. 
The results of the main paper has been obtained with embedded 8th order Runge-Kutta Prince-Dormand method with 9th order error estimate provided within the GNU Scientific Library.

\section{Fourier analysis of $R(t)$ in $\rm \bf tGA$}
 
\begin{center}
\begin{figure}[h!]
 \includegraphics[width=0.5\textwidth]{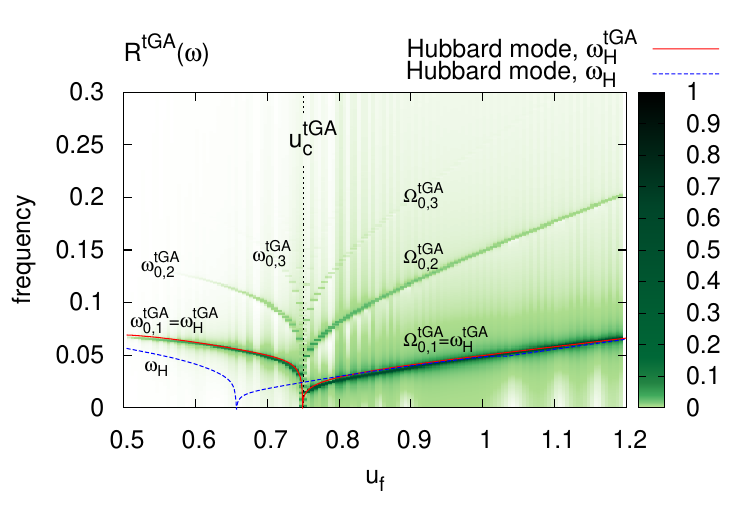} 
\caption{ Intensity (in arbitrary units) of frequencies obtained  from a Fourier analysis of the long-time evolution (${t\cdot8T_0\in\{0,1000\}}$) of the parameter $R$ (see main manuscript) under the time-dependent Gutzwiller approximation (tGA) \cite{Schiro2010S}.  The description of frequencies  with two integer numbers is analogical to this from main manuscript (cf. Eq. (13,14)) under assumption that $\omega_J=0$. For the comparison we also plot Hubbard mode resulting from the method presented in the main manuscript i.e., tGA supplemented with a variational time-dependent Schrieffer-Wolff transformation.}
\label{sfig2}
\end{figure}
\end{center}

Under the time-dependent Gutzwiller approximation (tGA) \cite{Schiro2010S} the dynamical properties of the quenched Hubbard model at half-filling can be studied by solving following coupled differential equations,
\begin{equation}
 \begin{split}
  i\dot\phi_0 &= \frac{u_f}{2}\phi_0 - \frac{1}{4}(\phi_0\phi_1^*+\phi_0^*\phi_1)\phi_1,\\
i\dot\phi_1 &= -\frac{1}{4}(\phi_0\phi_1^*+\phi_0^*\phi_1)\phi_0.
 \end{split}
\end{equation}
 In Fig. \ref{sfig2} we show a Fourier analysis of real time evolution of $R=\phi_0\phi_1^*+\phi_1\phi_0^*$.  
% \widetext

 \end{document}